\documentclass[nato,noid,numreferences]{crckbked}

\usepackage{graphicx}
\begin{document}
\begin{opening}
\title{COLOR CONFINEMENT AND DUAL SUPERCONDUCTIVITY: AN UPDATE}
\author{Adriano Di Giacomo}
\institute{Dipartimento di Fisica, Universit\`a di Pisa and\\
	INFN Sezione di Pisa\\
	Via Buonarroti 2, I-56127 Pisa, Italy}
%
\runningtitle{Color Confinement}
\runningauthor{A.\ DI\ GIACOMO}
\begin{abstract}
The evidence for dual superconductivity as a mechanism for
color confinement is reviewed. New developments are presented for full
QCD, i.e. in the presence of dynamical quarks.
\end{abstract}
\end{opening}
\section{Introduction}
Confinement is by definition the absence of free colored particles in
nature. In spite of the clear evidence coming from high energy
experiments that quarks and gluons are the fundamental constituents of
hadrons, none of them has ever been detected.

   Only upper limits exist to production cross sections .The cross
section $\sigma_q$ for the inclusive process
\begin{equation}
p + p \to q(\bar q) + X\label{eq1}
\end{equation}
has an upper limit\cite{1}
\begin{equation}
\sigma_q < 10^{-40}{\rm cm}^2\label{eq2}
\end{equation}
At the same energies the total cross section $\sigma_T$   has the value
\begin{equation}
\sigma_T\simeq 10^{-25}{\rm cm}^2\label{eq3}
\end{equation}
In perturbation theory the ratio $\sigma_q/\sigma_T$  is expected
to be a sizable fraction of unity. From (2) and (3)
\begin{equation}
\sigma_q/\sigma_T < 10^{-15}\label{eq4}
\end{equation}
   Relic quarks in nature have been hunted since they were first proposed
as fundamental bricks of matter\cite{2}.

Fourty years of Millikan-like experiments looking for fractionally
charged particles have produced as upper limit [1]
\begin{equation}
n_q/n_p < 10^{-27}\label{eq5}
\end{equation}
for the ratio of the abundance of quarks $n_q$  and that of nucleons
$n_p$ in nature. The limit  (5)  results from the analysis of
  $\sim$ 1g  of matter and no quarks found. In the absence of confinement the
Standard Cosmological Model predicts\cite{3}
\begin{equation}
n_q/n_p \simeq 10^{-12}\label{eq6}
\end{equation}
Again
\begin{equation}
\frac{\displaystyle (n_q/n_p)_{obs} }{\displaystyle (n_q/n_p)_{expected}}
< 10^{-15}\label{eq7}
\end{equation}
   The only natural explanation of the limits (3) and (7) is that
  $\sigma_q$, $n_q$ are exactly zero, or that confinement is an absolute
property based on symmetry [4].
Similar situations are e.g.
\begin{itemize}
\item[i)] the photon mass, which is experimentally bounded by the 
inverse radius
      of the solar system\cite{1}.The corresponding symmetry is gauge
     invariance.
\item[ii)] ordinary superconductivity. The upper limit of the resistivity of
     superconductors is many orders of magnitude smaller than that of any
     other material.The symmetry pattern behind that is the Higgs
     phenomenon\cite{5}.
\end{itemize}
\section{Virtual tests (lattice)}
    QCD at finite temperature can be simulated on the lattice. It is a well
known theorem that the partition function of a field system is equal to
the euclidean Feynman integral, with immagianary time ranging from  0  to
$1/T$, and periodic boundary conditions for bosons, antiperiodic for
fermions:
\begin{equation}
Z = \int[{\cal D}\varphi]\,\exp\left[ - \int_0^{1/T}\!dt\int d^3x
{\cal L}(\vec x,t)\right]
\label{eq8}
\end{equation}

   If the lattice size in the time direction is $N_T$  then
  \begin{equation}
T = \frac{1}{N_T a}
\label{eq9}
\end{equation}
  $a$ being the lattice spacing in physical units.  Renormalization group
gives, at large enough $\beta=2N/g^2$, $a
=\frac{1}{\Lambda_L}\exp(-\beta/b_0)$, where $ b_0 > 0$ by asymptotic
freedom, i.e.
\begin{equation}
T = \frac{\Lambda_L}{N_T} \exp(\beta/b_0)
\label{eq10}
\end{equation}
   Low  $T$  corresponds to large  $g^2$   (strong coupling or disorder in the
language of statistical mechanics ); high  $T$ to weak coupling or order.
This is the opposite to what happens in ordinary spin systems, for which
$T$  plays the role of  $g^2$.

In pure gauge theories an order parameter for confinement is 
$\langle L\rangle$,
the Polyakov line, which is the trace of the parallel transport along the
immaginary time axis from 0  to  $1/T$, closed by periodic
boundary conditions. The corresponding symmetry is  $Z_N$.

   On the lattice the correlator
\begin{equation}
G(\vec r) = \langle L(\vec r) L^\dagger(\vec 0)\rangle
\label{eq11}
\end{equation}
is measured\cite{6}.
    Cluster property requires
\begin{equation}
G(\vec r) \mathop\simeq_{r\to \infty} A\exp(-\sigma a N_T r) +
|\langle L\rangle |^2
\label{eq12}
\end{equation}
The potential energy of a static $q\bar q$ pair at distance $r$ is given by
\begin{equation}
V(r) = -\frac{1}{a N_T} \ln G(r)
\label{eq13}
\end{equation}
A temperature $T_c$ is found such that for $T < T_c$  $\langle L\rangle =0$  or
\begin{equation}
V(r) \mathop\simeq_{r\to \infty} \sigma r
\label{eq14}
\end{equation}
which means confinement. For $T > T_c$  $\langle L\rangle \neq 0$  and
\begin{equation}
V(r) \mathop\simeq_{r\to \infty} const.
\label{eq15}
\end{equation}
which means deconfinement.

   Finite size scaling analysis of the correlator around the critical point
provides a determination of the critical index $\nu$. For  $SU(2)$ pure
gauge theory the transition is second order\cite{6}, consistent with the class
of universality of the 3d ising model ($\nu  =.62$) as expected\cite{7}, and
$T_c/\sqrt{\sigma}\simeq .7$. For $SU(3)$ pure gauge theory the 
transition is weak
first order \cite{8,9}, ($\nu   =.33$) and  $T_c/\sqrt{\sigma}\simeq .65$,
which, by the usual assumption $\sqrt{\sigma} = 425\,{\rm MeV}$ gives
$T_c \simeq 270\,{\rm Mev}$.

In the presence of dynamical quarks $Z_N$ is explicitely broken , and 
$\langle L\rangle
$ cannot be an order parameter. For two equal-mass dynamical quarks the
situation is depicted in fig.1.

\begin{figure}[htb]
\centering
\includegraphics[width=8cm,angle=270]{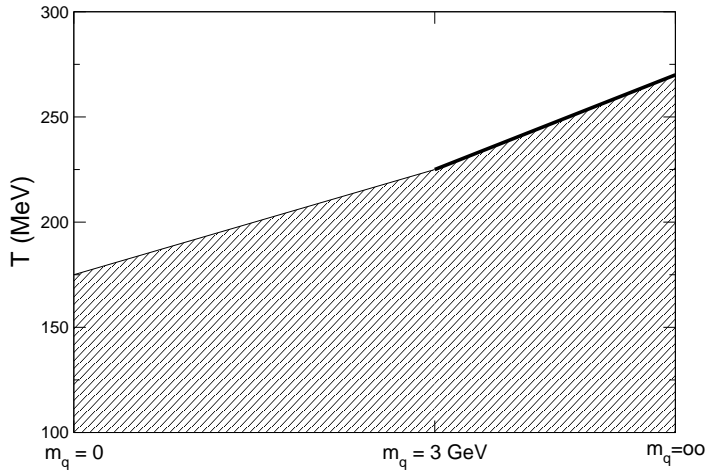}
\caption{Phase dyagram of QCD:shaded region is confined.}\label{Fig1}
\end{figure}

The transition temperature is
determined, at given quark mass, by looking at the maximum of a number
of suceptibilities , e.g.
$\int \langle\bar\psi\psi(x)\, \bar\psi\psi(0)\rangle d^3x$,
$\int \langle L(x) L(0)\rangle d^3 x$.
All of them show a maximum at the same $T_c$\cite{10}. For high enough
$m_q$, ($m_q \geq 3\, {\rm GeV}$) the maximum of the Polyakov line
susceptibility  goes large
   with increasing spatial volume as in the quenched case: a finite size
scaling analysis shows that the transition is first order. There are
indications that the transition is second order in the chiral limit
$m_q=0$,as suggested by symmetry arguments\cite{11}. At intermediate values
of $m_q$ none of the susceptibilities which have been considered
increases with increasing spatial volume, and a possible conclusion is
that there is no phase transition but only a crossover. The overall
situation is rather confusing. It is not clear a priori what is the
relation between chiral symmetry and confinement. It is not fully clear
either what susceptibilities are entitled to determine the order of the
transition by their behavior at large volumes. In principle the relevant
quantities should be those appearing in the expression of the free
energy. The free energy (effective lagrangean ) depends on the dominant
exitations and on their symmetry. What are the dominant exitations is
exactly the problem under investigation.
\section{Duality.}
   Confined phase is disordered. How can the symmetry of a disordered phase
be defined? The key concept is duality\cite{12}. It applies to d-dimensional
systems admitting non trivial topological excitations in (d-1) dimensions.

These systems admit two complementary descriptions.
\begin{itemize}
\item[1)] A direct description in terms of the fields $\phi$,with order
parameters  $\langle\phi\rangle$, in which the topological
configurations  $\mu$ are non local. This description is convenient in the weak
coupling regime ($g \ll 1$), i.e. in the ordered phase.
\item[2)]
A dual description in which the topological excitations $\mu$ become
local fields, and the original fields $\phi$ topological
configurations. The dual coupling $g_D$ is related to  $g$  as
$g_D\sim 1/g$. This
description is convenient in the disordered phase (strong coupling
regime). Its symmetry is described by  $\langle\mu\rangle$ (disorder 
parameter).
Duality maps the strong coupling regime of the direct description into the weak
coupling regime of the dual description.
\end{itemize}
   The prototype system for duality is the ising model\cite{13} where dual
excitations are kinks. Other examples are N=2 SUSY QCD\cite{14}, where the
dual excitations are monopoles; M string theories\cite{15}; 3-d XY model,
where dual excitations are abelian vortices\cite{16}; 3-d Heisenmerg 
magnet, with
2-d Weiss domains as dual excitations\cite{17}; compact U(1) gauge 
theory , where
dual excitations are monopoles\cite{18,19}.

In QCD the dual topological excitations have to be identified : as we
will see, however, information exists on their symmetry. Two original
proposals exist in the literature, which have been widely studied:
\begin{itemize}
  \item[a)] Monopoles\cite{20,21}. The idea is that vacuum acts as a dual
superconductor, which confines electric charges by Meissner effect, in
the same way as magnetic charges are confined in an ordinary
superconductor. Developments of this approch will be the subject of the
next sections.
\item[b)]
Vortices\cite{4}. The symmetry involved is $Z_N$.
\end{itemize}
In 2+1 dimensions a
conserved charge exists, the number of vortices minus the number of
antivortices, and vortives are described by a local field. In 3+1
dimensions a dual Wilson loop can be defined ('tHooft loop ) $B(C)$, in
connection with any closed path  C . The algebra which is obeyed by $B(C)$
and by the ordinary Wilson Loop  $W(C')$ is
\begin{equation}
B(C) W(C') = W(C') B(C) \exp\left(i n_{CC'} \frac{2\pi}{N_c}\right)
\label{eq16}
\end{equation}
where $n_{CC'}$ is the linking number of the two loops. From eq(16) it follows
that, if  $\langle W(C')\rangle$ obeys the area law
$\langle B(C)\rangle$ obeys the perimeter law,
and if  $\langle B(C)\rangle$ obeys the area law then
  $\langle W(C')\rangle$ obeys the perimeter
law. If we denote by
$\langle L\rangle$ the ordinary Wilson loop which wraps the lattice through
periodic b.c. in time (Polyakov loop) ,and  by
$\langle \tilde L\rangle$ the analogous dual
loop ('tHooft's line),then in the confined phase
$\langle L\rangle = 0$, $\langle \tilde L\rangle \neq 0$,
whilst in the deconfined phase  $\langle \tilde L\rangle = 0$,
$\langle  L\rangle \ne 0$.
$\langle \tilde L\rangle$ is a
disorder parameter for confinement.
    These relations have been tested on the lattice\cite{22,23}. The
corresponding symmetries $Z_N$  and  $\tilde Z_N$ are explicitely broken in the
presence of fermions.
\section{Monopoles.}
   Monopoles in non abelian gauge theories are always abelian (Dirac)
monopoles. This statement can be immediately checked by looking at the
field produced by a static configuration of colored matter at large
distances, by use of the familiar multipole expansion [24]. Monopoles 
are identified by
a constant diagonal matrix in the algebra, with integer or half-integer
values : they carry  N-1 abelian magnetic charges.
   The same physics emerges from the procedure known as abelian projection
[21]. We shall illustrate it for $SU(2)$: the general case\cite{25} is not
substantially different. Let $\vec \varphi(x)$ be any operator in the adjoint
representation, and  $\hat\varphi(x) = \vec\varphi(x)/|\vec\varphi(x)|$  its
direction in color space.
Define\cite{26}
\begin{equation}
F_{\mu\nu} = \hat\varphi\vec G_{\mu\nu} -
\frac{1}{g}\hat\varphi(D_\mu\hat\varphi\wedge D_\nu\hat\varphi)
\label{eq17}
\end{equation}
with
$\vec G_{\mu\nu} = \partial_\mu \vec A_\nu - \partial_\nu\vec A_\mu
+ g \vec A_\mu\wedge \vec A_\nu$ the field strength
  and  $D_\mu = \partial_\mu + g \vec A_\mu \wedge$   the
covariant derivative. Both terms in eq.(17) are color singlets and gauge
invariant: the combination is chosen to cancel bilinear terms $A_\mu A_\nu$.
Indeed one has identically:
\begin{equation}
F_{\mu\nu} =
\hat\varphi(\partial_\mu \vec A_\nu - \partial_\nu\vec A_\mu) -
\frac{1}{g}\hat\varphi(\partial_\mu\hat\varphi\wedge \partial_\nu\hat\varphi)
\label{eq18}
\end{equation}
  In a gauge in which  $\hat\varphi$  is constant, e.g.
$\hat\varphi = (0,0,1)$, $F_{\mu\nu}$  is
abelian :
\[ F_{\mu\nu} = \partial_\mu A^3_\nu -
\partial_\nu A^3_\mu\]
   A magnetic current ,  $j_\mu$, can be defined in terms of the dual
tensor   $F^*_{\mu\nu} = \frac{1}{2}\varepsilon_{\mu\nu\rho\sigma}
F_{\rho\sigma}$,
\[ j_\mu      =    \partial_\nu F^*_{\mu\nu}\]
  $j_\mu$ is identically
zero ( Bianchi identities ) in a non compact formulation of the 
theory. In a compact
formulation, like Lattice, $j^\mu$ can be non zero. In any case it is 
identically
conserved
\begin{equation}
\partial_\mu j^\mu = 0
\label{eq19}
\end{equation}
Magnetic charges are Dirac monopoles, obeying Dirac quantization condition
  $Q = n/2g$.  The corresponding  magnetic $U(1)$ symmetry can either
be realized a la Wigner, and then the Hilbert space consists of
superselected sectors with definite magnetic charge, or Higgs-broken, and
then the system behaves as a dual superconductor. If the ideas of ref's
\cite{20,21} are correct the expectation is that QCD vacuum behaves as a
dual superconductor (Higgs-broken phase) for $T < T_c$, and as a magnetic
superselected system for $T > T_c$. A disorder parameter should
discriminate between superconductor and normal.
   Such a parameter has been constructed\cite{27,28,29} as the v.e.v. 
$\langle\mu\rangle$
of an operator $\mu$ carrying magnetic charge. In fact  $\mu$ is a Dirac-like
operator\cite{30}, charged and gauge invariant\cite{18,33}. The construction
of $\mu$  is at the level of a theorem for compact $U(1)$\cite{18,33}. In non
abelian gauge theories it is undefined by terms ${\cal O}(a^2)$, $a$ being the
lattice spacing. like the abelian projection itself\cite{32,34}.
\section{Results}
The basic structure of $\mu$  is a translation of the field configuration
in the Schr\"odinger picture by a classical monopole configuration. In the
same way as
\begin{equation}
e^{ip a}|q\rangle = |q + a\rangle\label{eq20}
\end{equation}
defining
\[\mu(\vec x,t) =
\exp\left( i \int d^3\vec y \Pi(\vec y,t)\bar\varphi(\vec x - \vec y)\right)
\]
   with $\Pi(\vec y,t)$ the conjugate momentum to the field $\varphi(\vec y,t)$,
and $\bar\varphi(\vec y,t)$  the classical field configuration to be added
\begin{equation}
\mu(\vec x,t)|\varphi(\vec y,t)\rangle
=|\varphi(\vec y,t) + \bar\varphi(\vec x - \vec y)\rangle
\label{eq21}
\end{equation}
   In fact the basic structure has to be adapted to a compact formulation,
in which the field cannot be translated at will\cite{19}, and to the abelian
projected situation, in which only the abelian part of the field has to
be translated. All this has been done\cite{27}. The resulting disorder
parameter $\langle\mu\rangle$ can be finally written as the ratio of 
two partition
functions
\begin{equation}
\langle\mu\rangle = \frac{\displaystyle \tilde Z(\beta)}{
\displaystyle  Z(\beta)}
\label{eq22}
\end{equation}
with $Z(\beta) = \int[{\cal D}\varphi]\exp(-\beta S)$, and
\begin{equation}
Z(\beta) = \int[{\cal D}\varphi]\exp\left[-\beta (S + \Delta S)\right]
\label{eq23}
\end{equation}
$S + \Delta S$ is obtained from $S$ by a modification of the space time
plaquettes at time $t$, $\Pi_{0i}(\vec n,t)$; for details see ref's
\cite{19,27}. Instead of $\langle\mu\rangle$ itself it is more 
convenient to study
\begin{equation}
\rho = \frac{d}{d\beta}\ln\langle\mu\rangle =
\langle S\rangle_S - \langle S + \Delta S\rangle_{S\Delta S}
\label{eq24}
\end{equation}
On the one hand  $\rho$ is numerically easier, since 
$\langle\mu\rangle$ fluctuates
wildly as any partition function; on the other hand we shall see that $\rho$
contains all the relevant information as $\langle\mu\rangle$, and 
also in a more
convenient way. From $\rho$, $\langle\mu\rangle$  is obtained as
\begin{equation}
\langle\mu\rangle = \exp\left[\int_0^\beta \rho(x) dx\right]
\label{eq25}
\end{equation}
since  $Z(\beta=0) = \tilde Z(\beta=0) = 1$.

  The typical shapes of $\langle\mu\rangle$  and $\rho$  as  functions 
of $\beta$ are
plotted in fig.2. The position of the negative peak coincides with 
the deconfining
phase transition.

\begin{figure}[htb]
\centering
\includegraphics[width = 0.45\linewidth]{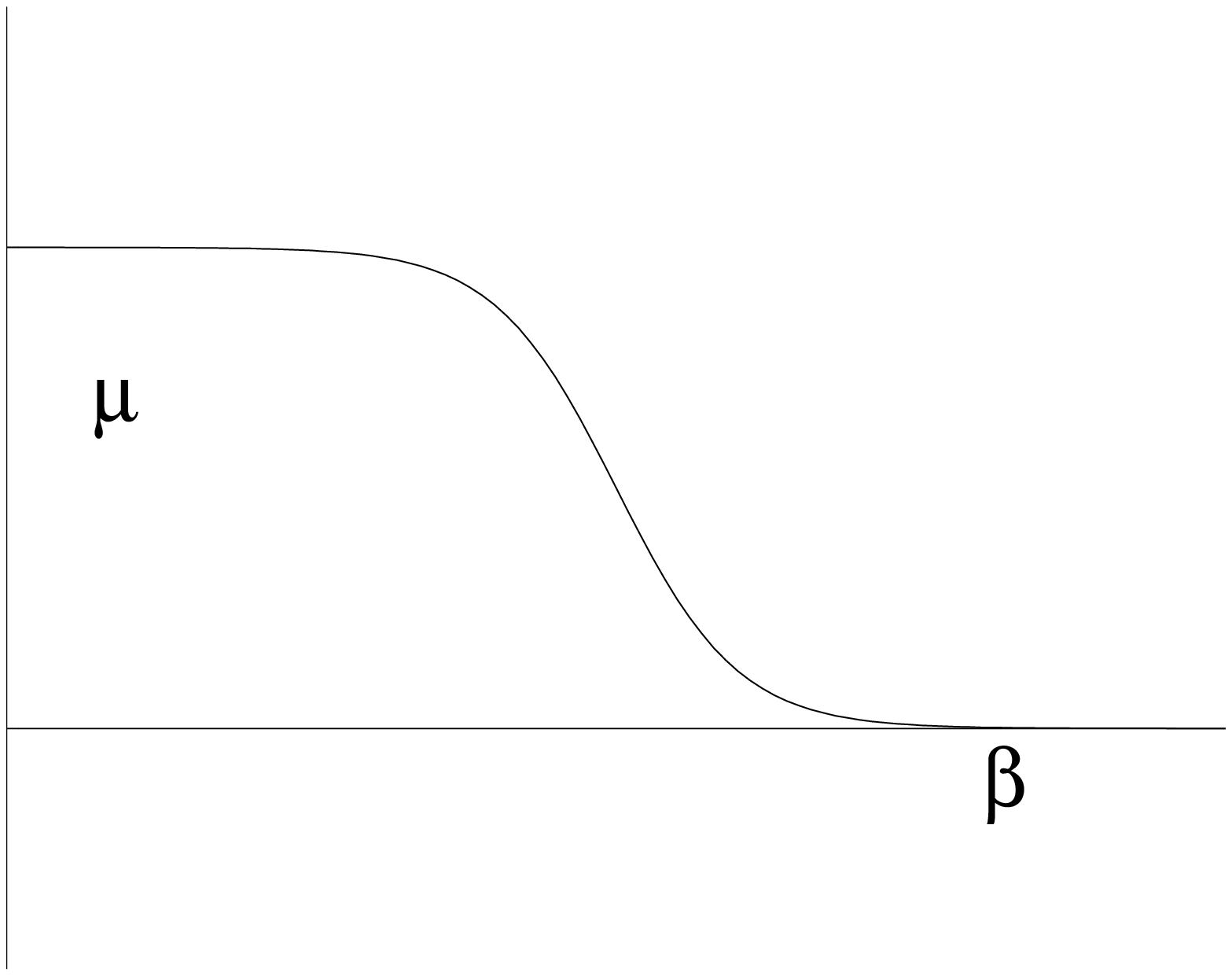}
\includegraphics[width = 0.45\linewidth]{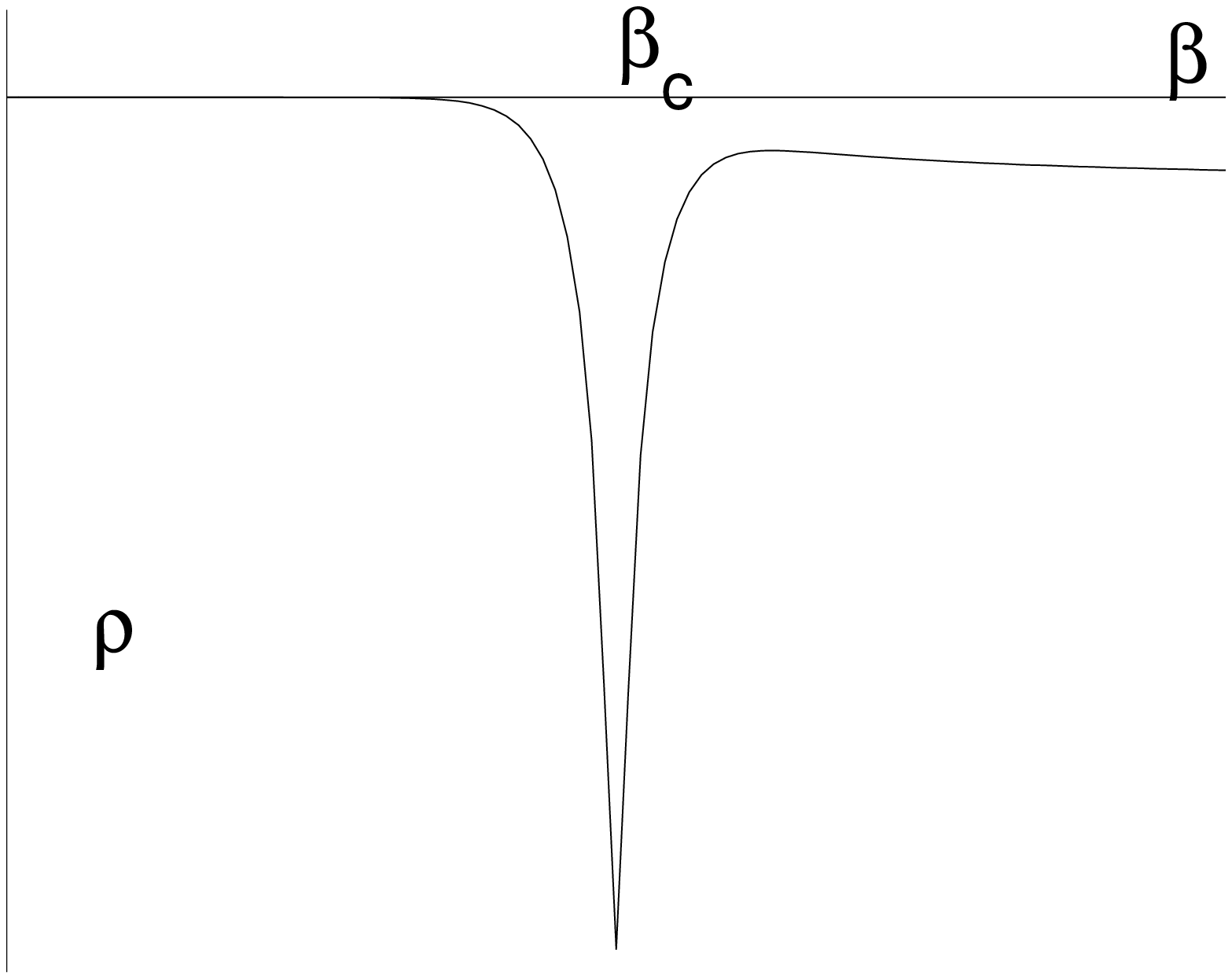}
\caption{Typical shape of $\langle\mu\rangle$ and $rho$.}\label{Fig2}
\end{figure}

Fig.3 shows $\rho$ for different spatial sizes  $N_s$ of the
lattice, at fixed $N_t=4$, for $SU(2)$ pure gauge theory.

\begin{figure}[htb]
\centering
\includegraphics[width=8cm,angle=270]{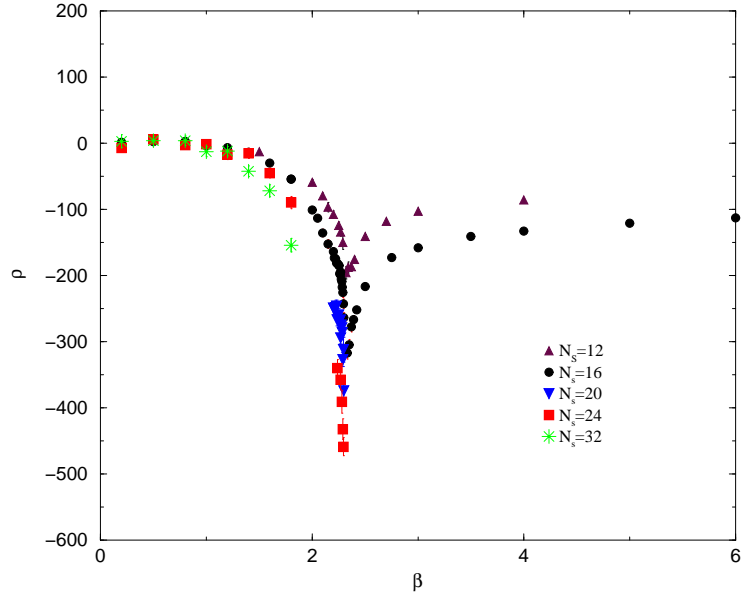}
\caption{$SU(2)$: $\rho$ at different spatial sizes.}\label{Fig3}
\end{figure}

The position of the
peak coincides with the maximum of the susceptibility of the Polyakov
line, as determined in ref\cite{6}, i.e. with the phase transition.
   In the range of temperatures  $T < T_c$ $\rho$ stays practically constant by
increasing the volume, as shown in fig.4 ,and this means that 
$\langle\mu\rangle$
has a non zero limit in that region.

\begin{figure}[htb]
\centering
\includegraphics[width=8cm,angle=270]{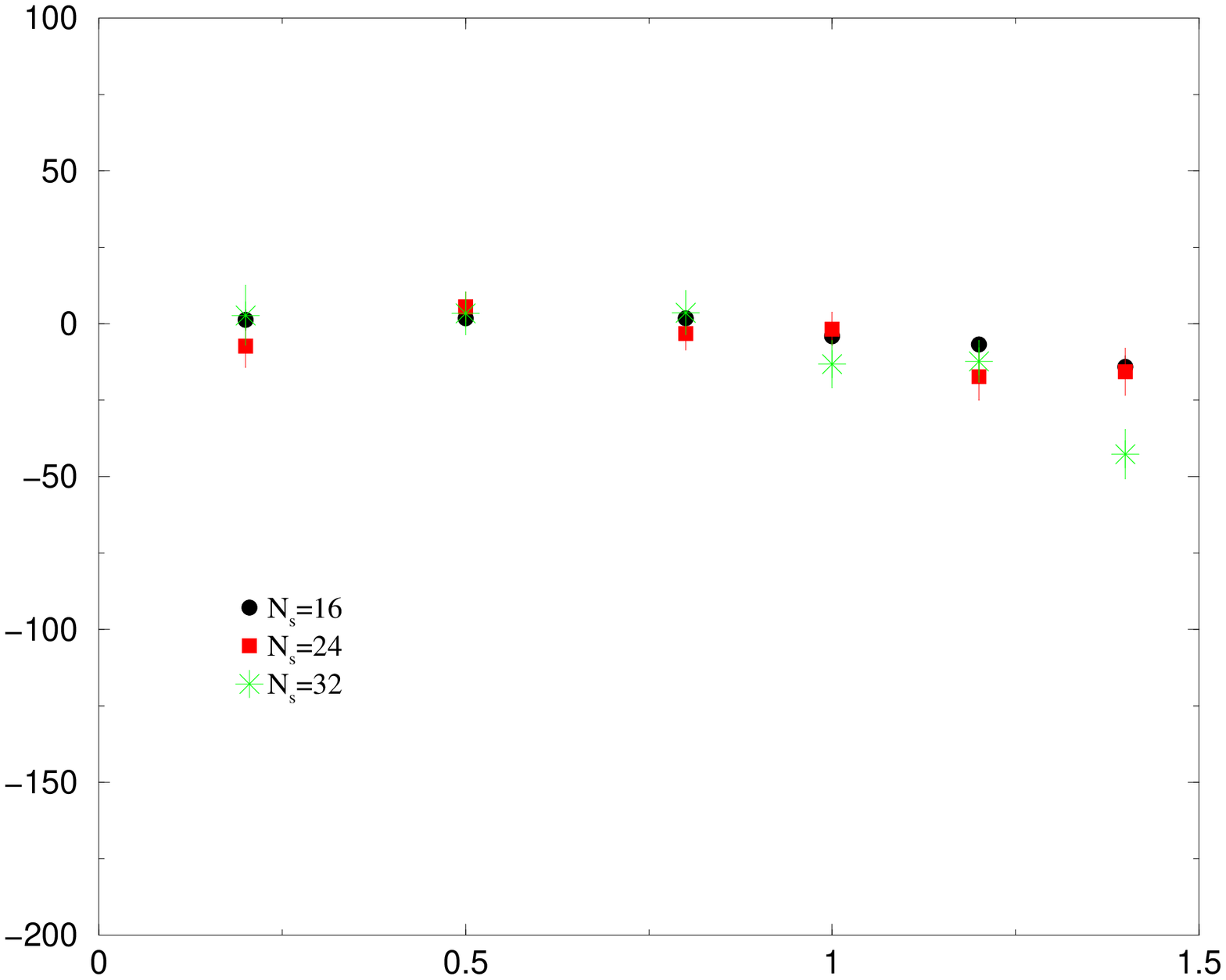}
\caption{$\rho$ at $T<T_c$ for different spatial sizes.}\label{Fig4}
\end{figure}

In the range $T > T_c$ $\rho$ diverges to $-\infty$ as $N_s$ goes
large, as
\begin{equation}
\rho = - k N_s + k'\qquad k > 0
\label{eq26}
\end{equation}
as shown in fig.5 .

\begin{figure}[htb]
\centering
\includegraphics[width=8cm]{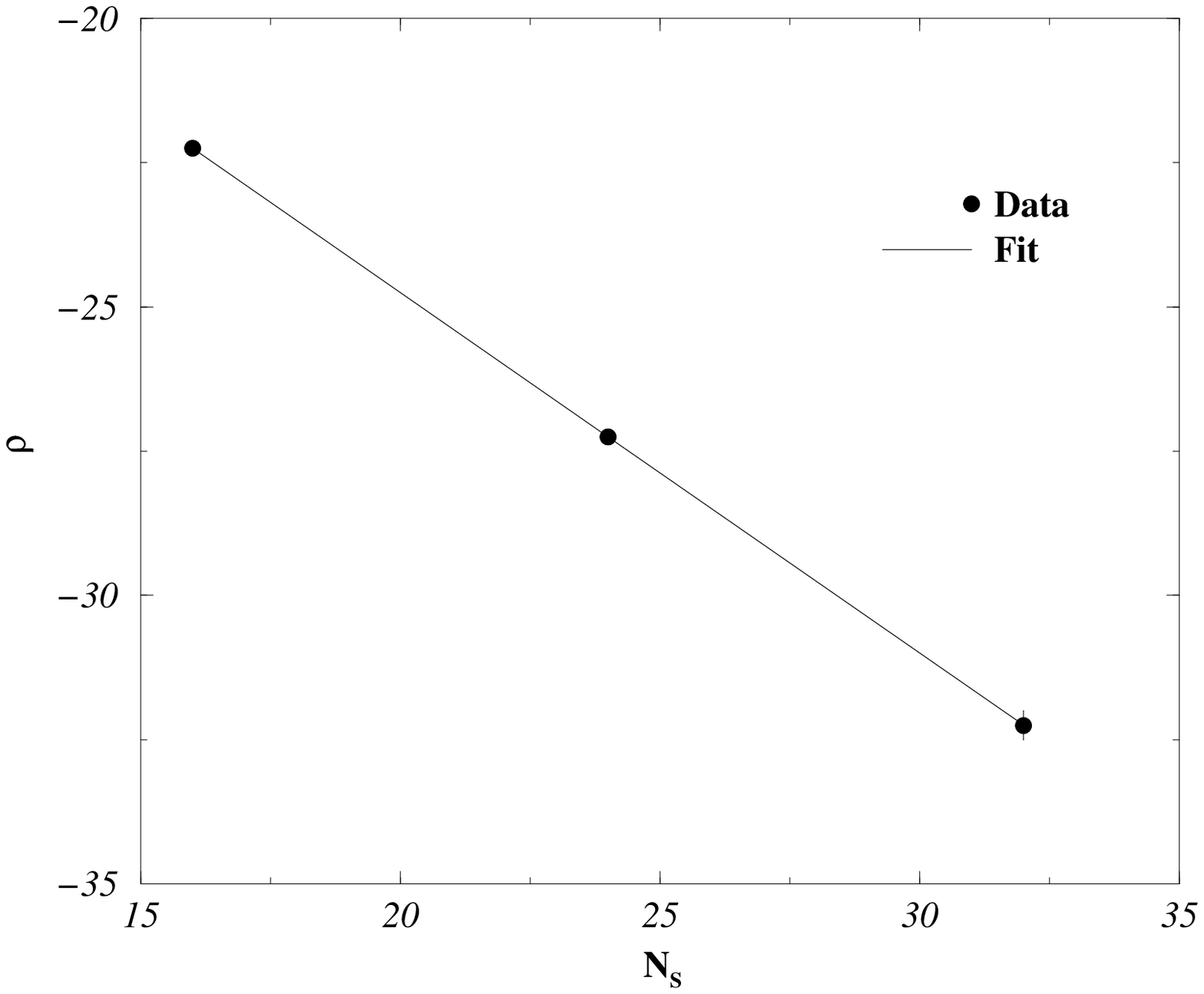}
\caption{$\rho$ at $T > T_c$ as a function of spatial size.}\label{Fig5}
\end{figure}

If we had measured $\langle\mu\rangle$ itself instead of
$\rho$  we
would have found it consistent with zero within large errors at large $ N_s$.
Measuring instead $\rho$ and checking numerically the behavior eq.(31) amounts
to state that $\langle\mu\rangle$ is exactly zero in the thermodynamical limit.
   In the critical region, around  $T_c$  one expects
\begin{equation}
\langle\mu\rangle \mathop\simeq_{T\to T_c}
\tau^\delta\Phi(\frac{a}{\xi},\frac{N_s a}{\xi})
\label{eq27}
\end{equation}
with  $\tau = 1 -\frac{T}{T_c} \propto (\beta_c - \beta)$, $\delta$  a critical
index,
$a$  the lattice spacing and $\xi$
      the correlation length.  The transition is known to be second order
for $SU(2)$, weak first order for $SU(3)$ gauge theory, and therefore
  $\xi\sim \tau^{-\nu}$ goes large at small  $\tau$'s. Neglecting 
$a/\xi\sim 0$ and
trading the variable $N_s/\xi$
with  $\tau N_s^{1/\nu}$, the scaling law follows for $\rho$   from
eq.(32)
\begin{equation}
\frac{\displaystyle\rho}{\displaystyle N_s^{1/\nu}} =
\frac{\displaystyle\delta}{\displaystyle\tau N_s^{1/\nu}}
+\Phi'(0,\tau N_s^{1/\nu})
\label{eq28}
\end{equation}
For different sizes of the lattice  $\rho/ N_s^{1/\nu}$, when plotted versus
$\tau N_s^{1/\nu}$
is expected to be independent of  $N_s$.
   A best fit to the data\cite{27,28,29} allows a determination of 
$\beta_c, \nu, \delta$.
The quality of the scaling is shown in fig.6  for $SU(2)$. The result is
\begin{eqnarray*}
SU(2)\qquad \nu = .62(1) && \delta = .20(3)\\
SU(3)\qquad \nu = .33(1) && \delta = .50(3)
\end{eqnarray*}

\begin{figure}[htb]
\centering
\includegraphics[width=8cm]{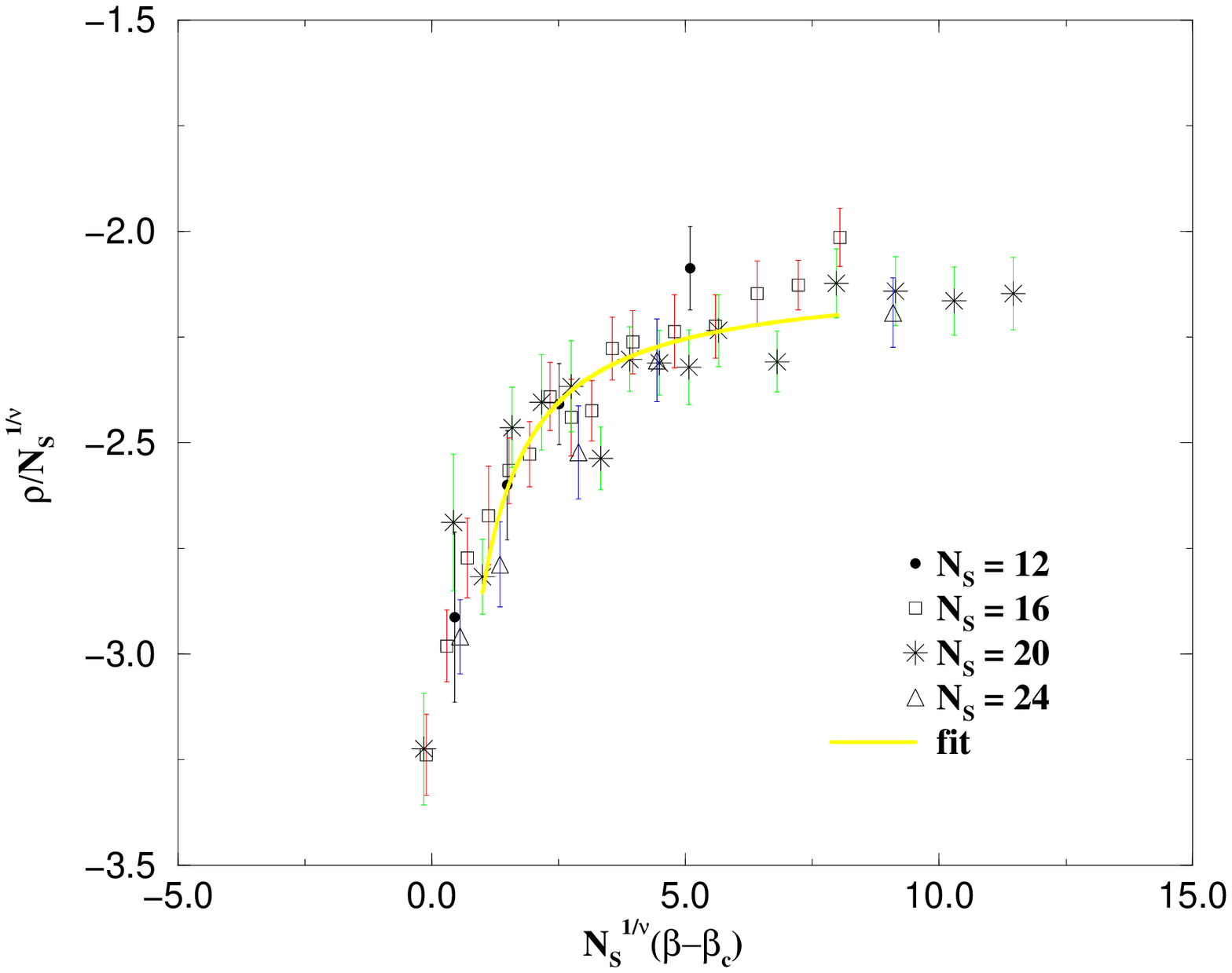}
\caption{Finite size scaling of $\rho$}\label{Fig6}
\end{figure}

   All that is for quenched theory. A few different abelian projections
have been tested, and the behavior of  $\rho$ ,as well as the value 
of  $\nu$  and
$\delta$  are independent of the abelian projection. An additional 
test has been made
with no abelian projection, i.e. by assuming as diagonal operators the
nominal $\lambda_3,\lambda_8$  used in the simulation\cite{29}. This
amounts to perform an average over all abelian projections, and the 
result does not
change.

   We can thus state that the confining phase of a pure gauge theory is a
dual superconductor in all the abelian projections, and undergoes a
transition to normal at $T_c$. What we learn from that is that, whatever the
dual excitations are, they carry magnetic charge in all the abelian
projections.
   The definition of the operator $\mu$  can  be easily extended to full QCD,
with dynamical quarks. A natural question is whether also in this case
the confining phase is characterized by dual superconductivity. This
would indeed be the expectation in the spirit of the  $N_c\to\infty$  limit.
If the physics of gauge theories is determined by the limit
$N_c\to\infty$ at
   $g^2 N_c$  fixed, corrections $1/N_c$ being small, then the mechanism of
confinement has to be  $N_c$  independent, and also insensitive to the
presence of dynamical quarks, their contribution being non leading in the
$1/N_c$ expansion.
    The parameter  $\rho = \frac{d}{d\beta}\ln\langle\mu\rangle$
  should go to  $-\infty$
for  $T > T_c$ in the thermodynamical limit so that 
$\langle\mu\rangle=0$: this is
indeed the case \cite{35}. For  $T < T_c$    $\rho$  converges to a
finite limit, i.e.  $\langle\mu\rangle\neq0$ and there is superconductivity
\cite{35}. The shaded area of fig.1 does indeed correspond to a 
superconductor, the
upper area to a superselected magnetic system: the negative peak is at the
transition as defined in sect.2 [fig.7]

\begin{figure}[htb]
\centering
\includegraphics[width=8cm]{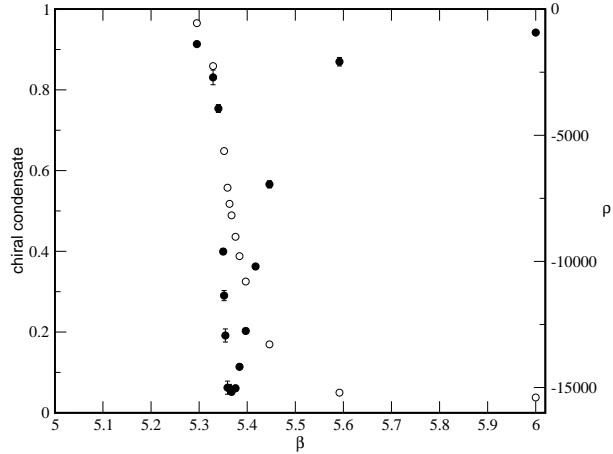}
\caption{2 flavour QCD. Open circles represent the chiral condensate, with
the scale axis on the left. Full circles represent $\rho$. The peak of $\rho$
coincides with the drop of $\langle\bar\psi\psi\rangle$.}
\label{Fig7}
\end{figure}

Around $T_c$ a finite size scaling analysis can give information on the
order of the transition : the difference with respect to the quenched case
is that now an extra dimensionful quantity, the quark mass, enters and
there is a two variable scaling. As usual we expect
\begin{equation}
\langle\mu\rangle \mathop\simeq_{T\to T_c}
\tau^\delta\Phi(\frac{a}{\xi},\frac{N_s a}{\xi}, m_q N_s^\alpha)
\label{eq34}
\end{equation}
In the critical regime  $a/\xi \simeq 0$ ,  $N_T/N_s\simeq 0$,
$\xi\sim \tau^{-\nu}$
and we can again
trade  $\xi/N_s$ with  $\tau N_s^{1/\nu}$, and the scaling law becomes
\begin{equation}
\langle\mu\rangle \mathop\simeq_{T\to T_c}
\tau^\delta\Phi(0,\frac{N_s a}{\xi}, m_q N_s^\alpha)
\label{eq35}
\end{equation}
   The index  $\alpha$  is known: one can chose for different values of  $N_s$
suitable quark masses, so as to keep $m_q N_s^\alpha$ fixed. The 
scaling law is then
the same as eq (28) ,whence the index $\nu$  can be extracted and with it
information on the order of the transition. This is a heavy numerical
program which is on the way on our APE machines. The result will possibly
be relevant to the problems discussed in sect 2.
\section{Conclusions and outlook.}
\begin{itemize}
\item[1)] Confinement is characterized by dual superconductivity of 
the vacuum in
all the abelian projections, both in quenched and in full QCD, in line
with the idea of the limit  $N_c\to\infty$.
   A universal disorder parameter has been defined for it.
\item[2)]  We do not know the dual excitations of  QCD  , nor their effective
lagrangean. However we know that they carry magnetic charge in all the
abelian projections.
\item[3)] An analysis of the critical region in full QCD is on the way ,which
will possibly clarify the nature of the phase transition depicted in fig.1.
\item[4)]
   The determination of the parameters of the dual supercondutor , Higgs
mass, penetration depht,\ldots is fundamental and relevant to understand the
deconfinement signsls which could come from heavy ion collisions: this
is also part of our research program.
\end{itemize}
   The contributions  of  L. Del Debbio, M. D'Elia , B. Lucini, G. Paffuti
to the research program are aknowledged.

\end{document}